\documentclass[useAMS,usenatbib]{mn2e}

\usepackage{graphicx}
\usepackage{times}
\usepackage{natbib}

\renewcommand{\d}{\mathrm{d}}

\title[Metals and molecules in structure formation]  
{Metal and molecule cooling in simulations of structure formation}
\author[U. Maio et al.]  
{U.~Maio$^{1}$\thanks{
E-mail:
maio@mpa-garching.mpg.de (UM);
kdolag@mpa-garching.mpg.de (KD);
ciardi@mpa-garching.mpg.de (BC);
torna@sissa.it (LT)
},
K.~Dolag$^{1}$$^\star$,
B.~Ciardi$^{1}$$^\star$,  
L.~Tornatore$^{2}$$^\star$\\
$^1$ Max-Planck-Institut f\"ur Astrophysik,
Karl-Schwarzschild-Strasse 1, D-85748 Garching b. M\"unchen, Germany\\
$^2$ SISSA/ISAS, Via Beirut 4, I-34014 Trieste, Italy\\
}

\begin{document}

\date{Accepted ???. Received ???; in original form April 2007}
\pagerange{\pageref{firstpage}--\pageref{lastpage}} \pubyear{2007}
\maketitle
\label{firstpage}

\begin{abstract}  
Cooling is the main process leading to the condensation of gas in the  
dark matter potential wells and consequently to star and structure  
formation. In a metal-free environment, the main  
available coolants are H, He, H$_2$ and HD; once the gas   
is enriched with metals, these also become important in defining the  
cooling properties of the gas.  
We discuss the implementation in Gadget-2 of molecular and metal cooling at  
temperatures lower that $\rm10^4~K$, following the time dependent  
properties of the gas and pollution from stellar evolution.  
We have checked the validity of our scheme comparing the results
of some test runs  
with previous calculations of cosmic abundance evolution and  
structure formation, finding excellent agreement.   
We have also investigated the  
relevance of molecule and metal cooling in some specific cases, finding that  
inclusion of HD cooling results in a higher clumping  
factor of the gas at high redshifts, 
while metal cooling at low temperatures can have a  
significant impact on the formation and evolution of cold objects.

\end{abstract}

  
  
\section{Introduction} \label{sect:intro}  
  
The understanding of cosmic structure formation and evolution is  
one of the most outstanding problems in astrophysics, which   
requires dealing with processes on very large  
scales, like galactic or cluster properties, and,  
at the same time, very small scales, like atomic behavior of gas and  
plasma. To join these extremes, it is fundamental to include atomic  
physics into astrophysics and cosmology.   
In fact, only with a unified study  
it was and is still possible to justify many physical phenomena  
otherwise not explained, like for example the very well known OIII  
forbidden line, typical of many gaseous nebulae.  
Interesting introductions into this subject are found in   
\cite{Spitzer1978}, the paper review by \cite{Osterbrock1988}   
and \cite{Osterbrock1989}.  
  
Nowadays, one of the main links between ``small scales'' and ``large  
scales'' seems to be the cooling properties of the gas, as, to  
form cosmic structures, it is necessary for the gas to  
condense in the dark matter potential wells and emit energy as  
radiation \citep[comprehensive reviews on the topic  
are][]{Barkana_Loeb_2001, ciardi2005}.   
For this reason it is fundamental to investigate the chemical  
properties of molecules and atoms and their cooling capabilities.  
In the standard cosmological scenario for structure formation, the  
first objects are supposed to form in metal-free halos with virial  
temperatures lower than $\rm 10^4~K$, for which atomic cooling is  
ineffective. In such physical conditions   
the most efficient coolants are likely to be molecules   
\citep[e.g.][]{LS1984, Puy_et_al_1993}.\\  
As hydrogen is the dominant element in the Universe, with a  
primordial mass fraction of about 76\%, we expect that the  
derived molecules will play a role in the cosmological gas chemistry.
The first studies in this direction were made by  
\cite{SaZi67} followed by \cite{Peebles_Dicke68} and many others  
\citep[][]{HM79,Abel_et_al1997,GP98,SLD_1998}, who
highlighted the importance of $\rm H_2$ in cooling gas   
down to temperatures of about $\rm 10^3~K$.\\  
In addition, one should also consider that, besides hydrogen,  
nucleosynthesis calculations predict the existence of  
primordial deuterium and lithium.
Recent measurements from a metal-poor damped  
Lyman $\alpha$ system \citep{Mit2006} give $\rm log(D/H)~=~-4.48~\pm~0.06$  
and are consistent with other observations   
\citep[][]{BurlesTytler98, PettiniBowen2001},  
while the abundance of Li (around $10^{-10}$)
is not very well determined and can vary by a factor of two or three  
when compared to the measurements in the atmospheres of old stars  
\citep[][]{Korn2006,Yong2006}. Other molecules derived from Li (e.g.   
LiH and $\rm LiH^+$) have much lower abundances   
\citep[e.g.][]{LS1984,Puy_et_al_1993,GP98}.
\\  
Another potentially interesting molecule is HD.   
Due to its permanent electric dipole moment\footnote{  
Some values of the permanent HD electric dipole moment  
found in the literature are   
$\rm D=8.3 \cdot 10^{-4}debye$ \citep{Abgralll1982} and  
$\rm D=~8.51~\cdot~10^{-4}debye$ \citep{Thorson1985}.
The first data date back to \cite{McKellar1976};  
for a theoretical, $ab\,initio$, non relativistic, perturbative  
treatment, via radial Schroedinger equation, see also   
\cite{Ford1977} and references therein.  
},  
HD has higher rotational transition probabilities  
and smaller rotational energy separations  
compared to $\rm H_2$ and thus,  
despite its low abundance of $\rm log(HD/H_2) \sim~ -4.5$  
\citep[e.g.][]{LS1984,Puy_et_al_1993,GP98}, 
HD can be an efficient coolant
\citep[][]{Flower2000,Galli_Palla_2002,HDcoolingfct,Abgrall2006}
and bring the gas in primordial halos to temperatures of the  
order of $\rm 10^2~K$.
This results into a smaller Jeans mass and a more efficient
fragmentation process.
For halos with virial temperatures in the range $\rm 10^3~K - 10^4~K$, HD  
cooling can be as relevant as H$_2$, while its effects are expected  
to be minor for larger halos \citep[see][]{Ripamonti2007,Shchekinov_Vasiliev_2006}.\\
The formation of primordial structures and stars have been 
investigated by many authors
\citep[like][]{BCL99, BCL02, Yoshida2006_2_astroph, Karlsson_2006}
but our understanding of the problem is still limited, because we are  
lacking informations on all those feedback effects (such as metal  
pollution, mass loss and energy deposition from the first stars) that  
profoundly affect it.
In particular, it is now commonly accepted that the  
presence of metals, by determining the cooling (and thus fragmentation)  
properties of a gas, influences the shape of the initial mass function (IMF),  
leading to a transition from a top-heavy IMF to a Salpeter-like IMF,  
when a critical metallicity --
varying between $\sim$$ 10^{-6}Z_{\odot}$ and
$\sim$$10^{-3.5}Z_{\odot}$, according to different authors
\citep[][]{Schneider2003,Bromm_Loeb_2003,Schneider_et_al_2006} --
is reached;
observationally, there are only few constraints
\citep[][]{Frebel_et_al_2007}.\\
\cite{Tornatore_et_al_2004} have presented the first  
implementation of a detailed chemical feedback model in the numerical
code Gadget \citep[other works on this subject are]
[]{Raiteri_et_al_1996, Gnedin98, KG2003, RicottiOstriker04}, through
which they study metal enrichment for different feedback/IMF scenarios.\\
In this study we discuss the implementation in Gadget of molecular and
metal cooling at temperatures below $10^4$~K and
we present a scheme able to deal both with  
primordial and metal enriched composition.
In details, we extend the existing implementation in Gadget of $\rm
H_2$ chemistry \citep{Yoshida_et_al2003},
in order to include HD, $\rm HeH^+$ and metal cooling
at those low temperatures. Indeed, these species are expected to be
relevant for the formation and evolution of cold objects.\\
The paper is organized in the following way:  
in Section \ref{sect:methods},   
we describe the computations of 
deuterium chemistry (Section \ref{sect:HD}),
metal lines (Section \ref{sect:met})
and their cooling capabilities (Section \ref{sect:cooling});
in Section \ref{sect:tests},   
we perform tests of our numerical implementation about   
chemical abundance evolution (Section \ref{sect:abundances}),  
cosmic structure formation (Section \ref{sect:large_scale})  
and cluster evolution (Section \ref{sect:cluster});  
in Section \ref{sect:conclusions},   
we discuss the results and give our conclusions.   
  
  
\section{Methods and tools} \label{sect:methods}  
  
In the commonly adopted scenario of structure formation, objects form  
from the collapse, shock and successive  
condensation of gas into clouds  
having a typical mass of the order of the Jeans  
mass. This process requires the gas to cool down, i.e. the  
conversion of kinetic energy into radiation that eventually escapes
from the system.  
This can occur via inelastic collisions which induce atomic electronic  
transitions to upper states, followed by de-excitations and   
emission of radiation.   
Details of the cooling process will depend on the type of elements considered  
and of transitions involved.\\  
In a standard primordial environment, the  
main coolants are expected to be hydrogen, helium and some molecules  
like $\rm H_2$ and HD; if the medium is metal enriched,   
the heavier elements become important coolants, thanks to a larger  
number of possible atomic transitions with different energy separations.\\   
The relevant quantity describing the cooling properties of a plasma is   
the energy emitted per unit time and volume,   
i.e. the cooling function (we will indicate  
it with $\Lambda$, adopting c.g.s. units,   
$\rm erg\, \rm cm^{-3}\, s^{-1}$)\footnote{  
Sometimes it is possible to find the same   
notation $\Lambda$ for the cooling rate in $\rm erg\, cm^{3}\, s^{-1}  
$.  
}.\\  
The characteristic time scale for the cooling, determined by  
$\Lambda$, is important to discriminate  
whether the gas can cool during the in-fall phase in the  
dark matter gravitational potential wells:  
structures are able to form only if the cooling time is short enough  
compared to the free-fall time.\\  
In the present work, we focus on the effects of molecules and metals  
in gas at low temperatures.  
In particular, we will include their treatment in  
Gadget-2 \citep{Springel2001, Springel2005}.  
This code uses a tree-particle-mesh algorithm to compute the  
gravitational forces and implements a   
smoothed-particle-hydrodynamics (SPH) algorithm to treat the baryons.  
Moreover, it is possible to follow the main non-equilibrium reactions   
involving electrons, hydrogen, helium and $\rm H_2$, with the respective  
ionization states \citep[][]{Yoshida_et_al2003}.   
Stellar feedback processes and metal release from SNII, SNIa and  
AGB stars are also included together with metal  
cooling at temperatures higher than $\rm 10^4$~K  
\citep[for a detailed discussion see][and references  
therein]{Tornatore_et_al_2007}.\\  
In the following, we are going to discuss in detail our HD 
and metal line treatment at $T<10^4$~K.

\subsection{HD treatment} \label{sect:HD}  
The HD molecule primarily forms through reactions between primordial  
deuterium and hydrogen atoms or molecules:   
a complete model for the evolution of HD involves 18 reactions  
\citep{Nakamura_Umemura_2002}, but, as their solution becomes quite
computationally expensive when implemented in cosmological
simulations, we use only the set of reactions selected by
\cite{Galli_Palla_2002}, which are the most relevant for HD
evolution:
\begin{eqnarray}  
\rm	 H_2  +  D\phantom{^+}    &\rightarrow &\rm HD + H 	\label{kHD1} \\  
\rm 	 H_2  +  D^+ &\rightarrow &\rm HD + H^+ 	\label{kHD2}   
\end{eqnarray}  
which lead to HD formation;  
\begin{eqnarray}  
\rm	HD + H\phantom{^+}    &\rightarrow &\rm D + H_2 		\label{kHD3} \\  
\rm	HD + H^+  &\rightarrow &\rm D^+ + H_2 	\label{kHD4}   
\end{eqnarray}  
for HD dissociation and H$_2$ formation; and   
\begin{eqnarray}  
\rm	H^+ + D\phantom{^+}  & \rightarrow &\rm H + D^+		\label{kHD5} \\  
\rm	H\phantom{^+} + D^+  & \rightarrow &\rm H^+ + D		\label{kHD6}   
\end{eqnarray}  
for charge exchange reactions.\\
From reactions (\ref{kHD1}) - (\ref{kHD6}), we see that HD abundance   
primarily depends on the amount of primordial deuterium and on the  
H$_2$ fraction.  
  
\begin{figure}  
\begin{center}  
\includegraphics[width=0.45\textwidth]{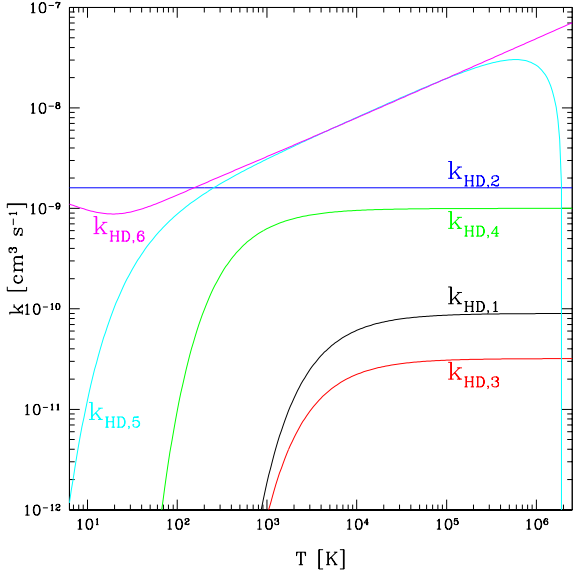}  
\caption{Temperature evolution of the reaction rates for deuterium  
chemistry. The labels refer to the number of the equation in the text.}  
\label{fig:rates}  
\end{center}  
\end{figure}  
  
For each species $i$, the variation in time of its number density $n_i$ is  
  
\begin{equation}\label{noneq_eq}  
\frac{\d n_i}{\d t}= \sum_p\sum_q k_{pq,i} n_p n_q - \sum_l k_{li} n_l  
n_i,  
\end{equation}  
where $k_{pq,i}$ in the first term on the right-hand side is the creation  
rate from species $p$ and $q$, and $k_{li}$ is the destruction rate  
from interactions of the species $i$ with the species $l$;  
they are temperature dependent and are usually expressed in $\rm  
[cm^3\, s^{-1}]$.\\  
A plot of the rates as a function  
of the temperature is given in Figure \ref{fig:rates}, and the exact  
expressions and references in Appendix \ref{app1}.  
From the figure, it is clear that the most important reactions in the  
 relevant range of temperatures are (\ref{kHD5}) and (\ref{kHD6}),  
and that the HD creation rates of reactions (\ref{kHD1}) and  
 (\ref{kHD2}) are always higher than the corresponding destruction  
 rates of reactions (\ref{kHD3}) and  (\ref{kHD4}), respectively.  
  
We have also considered $\rm HeH^+$ molecule evolution and found  
negligible effects on the cooling properties of the gas.   
The rates for $\rm HeH^+$ formation and destruction are given in  
Appendix \ref{app1}.\\
  
We implement our chemistry model extending the code by  
\cite{Yoshida_et_al2003}, which adopts the rates from  
 \cite{Abel_et_al1997}, and modify it for self-consistency to obtain a  
set of reactions including  
e$^-$, H, H$^+$, He, He$^+$, He$^{++}$, H$_2$, H$^+_2$, H$^-$,  
D, D$^+$, HD, $\rm HeH^+$ (a complete list of the reactions is given in Table  
\ref{tab:reactions}).\\  
The set of differential equations (\ref{noneq_eq}) is evaluated via  
simple linearization, according to a backward difference formula  
\citep{Anninos1997}:   
given the time step $\Delta t$, at each time $t$ and for each species  
$i$, equation (\ref{noneq_eq}) can be re-written as  
\begin{equation}\label{discrete}  
\frac{n_i^{t+\Delta t} - n_i^{t}}{\Delta t} = C_i^{t+\Delta t} -  
 D_i^{t+\Delta t} n_i^{t+\Delta t}  
\end{equation}  
where we have introduced the creation coefficient for the species $i$,  
in $\rm [cm^{-3} s^{-1}]$, as  
  
\begin{equation}  
C_i = \sum_p \sum_q  k_{pq,i} n_p n_q  
\end{equation}  
and the destruction coefficient, in $\rm [s^{-1}]$, as  
  
\begin{equation}  
D_i = \sum_l k_{li} n_l.  
\end{equation}  
The number density, $n_i^t$, is then updated from equation (\ref{discrete}):  
\begin{equation}  
n_i^{t+\Delta t} = \frac{C_i^{t+\Delta t} \Delta t + n_i^t}{1 +  
D_i^{t+\Delta t}\Delta t }.  
\end{equation}  
We apply this treatment to all chemical species.  
  
\begin{table}  
\begin{center}  
\caption{Set of reactions in the code.}\label{tab:reactions}  
\begin{tabular}{lr}  
\hline  
\hline  
Reactions & References for the rate coefficients\\  
\hline  
 	H    + e$^-$   $\rightarrow$ H$^{+}$  + 2e$^-$ & A97 / Y06\\  
	H$^+$   + e$^-$  $\rightarrow$ H     + $\gamma$    & A97 / Y06\\  
	He   + e$^-$   $\rightarrow$ He$^+$  + 2e$^-$    & A97 / Y06\\  
	He$^+$  + e$^-$   $\rightarrow$ He   + $\gamma$     & A97 / Y06\\  
	He$^+$  + e$^-$   $\rightarrow$ He$^{++}$ + 2e$^-$    & A97 / Y06\\  
	He$^{++}$ + e$^-$   $\rightarrow$ He$^+$  + $\gamma$ & A97 / Y06\\  
	H    + e$^-$   $\rightarrow$ H$^-$    + $\gamma$     & A97 / Y06\\  
	H$^-$    + H  $\rightarrow$ H$_2$  + e$^-$       & A97 / Y06\\  
        H    + H$^+$ $\rightarrow$ H$_2$$^+$  + $\gamma$ & A97 / Y06\\  
        H$_2$$^+$  + H  $\rightarrow$ H$_2$  + H$^+$     & A97 / Y06\\  
	H$_2$   + H   $\rightarrow$ 3H            & A97\\  
	H$_2$   + H$^+$ $\rightarrow$ H$_2$$^+$  + H     & S04 / Y06\\  
  	H$_2$   + e$^-$   $\rightarrow$ 2H   + e$^-$ & ST99 / GB03 / Y06\\  
      	H$^-$    + e$^-$   $\rightarrow$ H    + 2e$^-$   &A97 / Y06\\  
       	H$^-$    + H   $\rightarrow$ 2H    + e$^-$       & A97 / Y06\\  
       	H$^-$    + H$^+$ $\rightarrow$ 2H                &P71 / GP98 / Y06\\  
       	H$^-$    + H$^+$ $\rightarrow$ H$_2$$^+$  + e$^-$& SK87 / Y06\\  
        H$_2$$^+$  + e$^-$   $\rightarrow$ 2H            &GP98 / Y06\\  
        H$_2$$^+$  + H$^-$  $\rightarrow$ H    + H$_2$   &A97 / Y06\\  
        D    + H$_2$   $\rightarrow$   HD   + H     & WS02\\  
        D$^+$  + H$_2$   $\rightarrow$   HD   + H$^+$  & WS02\\  
        HD   + H   $\rightarrow$   D   + H$_2$         & SLP98\\  
        HD   + H$^+$  $\rightarrow$   D$^+$  + H$_2$   & SLP98\\  
        H$^+$  + D   $\rightarrow$   H    + D$^+$   & S02\\  
        H    + D$^+$  $\rightarrow$   H$^+$  + D    & S02\\  
        He    + H$^+$  $\rightarrow$   HeH$^+$  + $\gamma$    & RD82, GP98\\  
        HeH$^+$    + H $\rightarrow$   He  + H$_2^+$    & KAH79, GP98\\  
        HeH$^+$    + $\gamma$ $\rightarrow$   He  + H$^+$    & RD82, GP98\\  
\hline  
\hline  
\end{tabular}  
\end{center}  
Notes -  
P71~=~\cite{Peterson1971};  
KAH79~=~\cite{KAH1979};  
RD82~=~\cite{RD1982};  
SK87~=~\cite{SK1987};  
A97~=~\cite{Abel_et_al1997};  
GP98~=~\cite{GP98};  
SLP98~=~\cite{SLD_1998};  
ST99~=~\cite{ST99};  
WS02~=~\cite{Wang_Stancil_2002};  
S02~=~\cite{Savin_2002};  
GB03~=~\cite{GB03};  
S04~=~\cite{Savin_et_al2004};  
Y06~=~\cite{Yoshida2006_astroph}.  
\end{table}
  

\subsection{Metal treatment at $\rm T < 10^4~K$} \label{sect:met}  
For our calculations, we consider oxygen, carbon, silicon and iron,  
because they are the most abundant heavy atoms released during
stellar evolution and, therefore, they play the most important role in  
chemical enrichment and cooling: 
indeed, supernovae type II (SNII) expel mostly oxygen and carbon,
while supernovae type Ia (SNIa) silicon and iron 
\citep{Thielemann_et_al_2001,Ox_richSN_2003,SNejecta2004,Meynet_et_al_2006}.\\
We make the common assumption that carbon, silicon and iron are completely 
ionized, while oxygen is neutral. This is justified because, in a cosmological
context, UV radiation below 13.6~eV (from various astrophysical
sources, like quasars, stars, etc.) can escape absorption by neutral
hydrogen and generate a UV background that can ionize atoms with first
ionization potential lower than 13.6~eV
(like carbon, silicon and iron).
while oxygen remains predominantly neutral since its first ionization
potential of 13.62~eV is higher 
\citep[see also][]{Bromm_Loeb_2003,Santoro_Shull_2006}.\\

As in the low density regime of interest here thermodynamic equilibrium
is never reached (see discussion of eq.~\ref{n_cr}), the Boltzmann 
distribution for the population of atomic levels can not be used. Thus,
we will use the detailed balancing principle instead.
For each level $i$ of a given species, we impose that  
the number of transitions to that level (which $populate$ it),  
per unit time and volume 
equals the number of transitions from the same level $i$ to other  
levels (which $de-populate$ it), per unit time and volume:  
  
\begin{equation}\label{det_bal}  
n_i\sum_{j} P_{ij} = \sum_{j} n_j P_{ji}  \quad (i\neq j).  
\end{equation}  
In formula (\ref{det_bal}),   
$P_{ij}$ is the probability per unit time   
of the transition $i\rightarrow j$ and   
$n_i$ and $n_j$ are the number densities of atoms in the $i$-th and  
$j$-th (with $i \neq j$) level.  
The left-hand side of the previous equation refers to de-populations of  
the $i$-th level, while the right-hand side refers to the  
transitions which can populate it.\\
The probability of a given transition can be easily computed once   
the Einstein coefficients and the collisional rates are known.\\  
The further constraint which must be satisfied is the number particle  
conservation:
  
\begin{equation}\label{part_cons}  
\sum_{j} n_j = n_{tot}  
\end{equation}  
where $n_{tot}$ is the total number density of the species  
considered and $n_j$ the population of the generic level $j$.\\  
  
In case of collisional events, the rate at which the transition  
$i\rightarrow j$ occurs is by definition:  
\begin{equation}  
n_i n_x \gamma_{ij}\equiv  n_i n_x \langle u\sigma_{ij} \rangle =n_i n_x \int_0^\infty u \sigma_{ij} f(u)d^3u  
\end{equation}  
where $\sigma_{ij}$ is the cross section of the process, $f(u)d^3u$ is  
the velocity distribution function of the particles (typically a  
Maxwellian), $\gamma_{ij}$ is the collisional rate, 
$n_i$ the number density of the particles in the  
$i$-th level and $n_x$ is the colliding particle number density.\\  
The relation between $\gamma_{ij}$ and $\gamma_{ji}$ is:  
  
\begin{equation}  
g_{i}\gamma_{ij} = g_{j}\gamma_{ji} e^{-\beta \Delta E_{ji}} ,  
\end{equation}  
where $g_i$ and $g_j$  are the level multiplicities,  
$\beta=(k_B T)^{-1}$,   
$\Delta E_{ji}$ is the energy level separation and $i < j$.\\  
In addition to collisionally induced transitions, spontaneous  
transitions can take place with an emission rate given by the Einstein  
A coefficient.\\  
It is convenient to define the critical number density for the  
transition $ i\rightarrow j$ as  
  
\begin{equation} \label{n_cr}  
n_{cr,ij} = \frac{A_{ij}}{\gamma_{ij}}.  
\end{equation}  
This determines the minimum density above which  
thermal equilibrium can be assumed and low density deviations from  
the Boltzmann distribution become irrelevant. At densities below  
$n_{cr,ij}$, we expect values of the excited level populations lower than  
in the thermodynamic limit, because of the reduced number of  
interactions\footnote{The critical number density depends on the  
particular line transition considered; typical values for the fine  
structure transitions we are mostly interested in   
are of the order $\rm \sim 10^5 cm^{-3}$.  
}.\\
  
For a two-level system, the low density level  
populations arising from electron and hydrogen impact excitations can  
be found by solving the system of equations 
resulting from conditions 
(\ref{part_cons}) and (\ref{det_bal}):

\begin{equation}\label{sist}  
\left\{  
      \begin{array}{l}  
	\!\!\!\!n_1 + n_2 = n_{tot} \\  
	\\  
	\!\!\!\!n_1n_H\gamma_{12}^{H} \!+\!  
	n_1n_e\gamma_{12}^{e} \!-\!  
      	n_2n_H\gamma_{21}^{H} \!-\!  
	n_2 n_e\gamma_{21}^{e}\!-\!  
	n_2 A_{21} \!= \! 0  
     \end{array}  
     \right.  
\end{equation}  
where $n_H$ and $n_e$ are the hydrogen and electron number density,  
while $\gamma_{12}^{H}$  and  $\gamma_{12}^{e}$  
are the H-impact and e-impact excitation rate.\\  
The solution of (\ref{sist}) is:  
  
\begin{eqnarray}  
\frac{n_1}{n_{tot}}= \frac{\gamma_{21}^{H} + \gamma_{21}^{e}n_e/n_H +  
A_{21}/n_H}{\gamma_{12}^{H}+\gamma_{21}^{H}+\left(\gamma_{12}^{e}+\gamma_{21}^{e}\right)n_e/n_H  
+ A_{21}/n_H}\label{2lev_sol1}\\  
\frac{n_2}{n_{tot}}= \frac{\gamma_{12}^{H} +  
\gamma_{12}^{e}n_e/n_H}{\gamma_{12}^{H}+\gamma_{21}^{H}+\left(\gamma_{12}^{e}+\gamma_{21}^{e}\right)n_e/n_H  
+ A_{21}/n_H}\label{2lev_sol2} .  
\end{eqnarray}  
The ratio between the two level populations  
  
\begin{eqnarray}\label{n2_n1}  
\frac {n_2}{n_1} & = &  \frac{\gamma_{12}^{H} +  
\gamma_{12}^{e}n_e/n_H } {\gamma_{21}^{H} +  
\gamma_{21}^{e}n_e/n_H + A_{21}/n_H}\\  
& & \stackrel{n_e\ll n_H}{\sim}   
 \frac{\gamma_{12}^{H}}{\gamma_{21}^{H} + A_{21}/n_H}\label{n2_n1_ld}  
\end{eqnarray}  
will in general deviate from the Boltzmann statistic, because  
the spontaneous emission term dominates over the collisional term at low  
densities.  
In a neutral dense gas, instead, the level population saturates  
and simply reduces to a Boltzmann distribution, independently from the  
colliding particle number density.\\  
  
In case of $n-$level systems, one must solve the $n\times n$ population  
 matrix consisting of $n-1$ independent balancing equations  
 (\ref{det_bal}) and the constraint of particle conservation  
 (\ref{part_cons}).\\  
In the modeling, we approximate CII and SiII as a  
two-level system, and OI and FeII as a five-level system  
\citep{Santoro_Shull_2006}.\\  
Further details on the atomic data and structures are reported in  
Appendix \ref{app2}.  
  
\subsection{Cooling}\label{sect:cooling}  
  
\begin{figure}  
\begin{center}  
\includegraphics[width=0.45\textwidth]{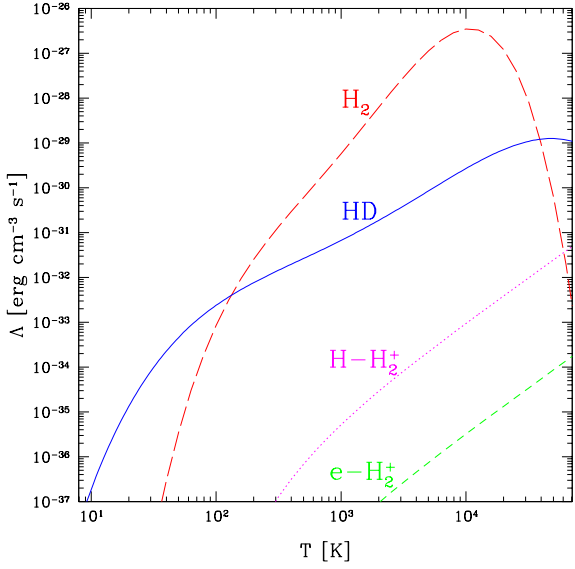}  
\caption{
Cooling functions for a primordial gas with a hydrogen number  
density of $\rm 1\, cm^{-3}$ and the following fractions for the  
different species:  
x$_{\rm HD} = 10^{-8}$,   
x$_{H_2} = 10^{-5}$,  
x$_{H_2^+} = 10^{-13}$, x$_{\rm e^-}= 10^{-4}$.  
The $\rm H_2$ cooling function (long-dashed line) is plotted together  
with the HD (solid), H-impact $\rm H_2^+$ (dotted line) and e-impact  
$\rm H_2^+$ (short-dashed line) cooling functions.  
 The fits for $\rm H_2^+$ cooling are appropriate only for $T <
10^4$~K.
}
\label{fig:HD_H2_H2II_cooling}  
\end{center}  
\end{figure}

In addition to calculating the chemical evolution of the gas, we need  
to evaluate the cooling induced by different species.  
In the original code, hydrogen and helium  
cooling from collisional ionization, excitation and recombination  
\citep[][]{Hui_Gnedin_1997},  
Compton cooling/heating and Bremsstrahlung  
\citep[][]{Black_1981} are evaluated.   
For the H$_2$ and H$_2^+$ cooling, the rates quoted   
in \cite{GP98} are adopted. We take the HD cooling function from  
\cite{HDcoolingfct}, who consider the HD ro-vibrational structure  
and perform calculations for $J \leq 8$ rotational levels and  
$v=0,1,2,3$ vibrational levels.  
Their results are somehow more accurate than other  
approximations \citep{Flower2000, Galli_Palla_2002} and valid for a  
wide range of number densities  
(up to $\rm 10^4\, cm^{-3}$) and temperatures ($\rm 10^2~K - 2\cdot
10^4 ~K$).\\  
In Figure \ref{fig:HD_H2_H2II_cooling}, we show cooling functions for  
H$_2$, HD, H$_2^+$ molecules; for the latter case we distinguish between  
neutral hydrogen impact and electron impact cooling; we have assumed  
fractions   
x$_{\rm HD} = 10^{-8}$,   
x$_{\rm H_2} = 10^{-5}$,  
x$_{\rm H_2^+} = 10^{-13}$,  
x$_{\rm e^-}= 10^{-4}$  
and a total hydrogen number density $\rm n_H = 1\, cm^{-3}$. Due to its  
very low abundance, H$_2^+$ is less effective than   
neutral H$_2$ and HD, which remain the only relevant coolants over the  
plotted range of temperature.\\  
We highlight that the contribution of HD to gas cooling at low  
temperatures is dominant in the case considered here, but its relevance  
strongly depends on the relative abundances of the species.\\  
  
The cooling for metal line transitions is computed as follows.  
In case of two-level systems, we define  
\begin{equation}\label{cooling_fct_def}  
\Lambda \equiv n_2 A_{21} \Delta E_{21}  
\end{equation}  
where $n_2$ is the atomic excited state number density,  
$A_{21}$ is the probability per unit time of the transition  
$2\rightarrow 1$ and $\Delta E_{21}$ is the energy separation of the  
levels.  
  
\begin{figure}  
\begin{center}  
\includegraphics[width=0.45\textwidth]{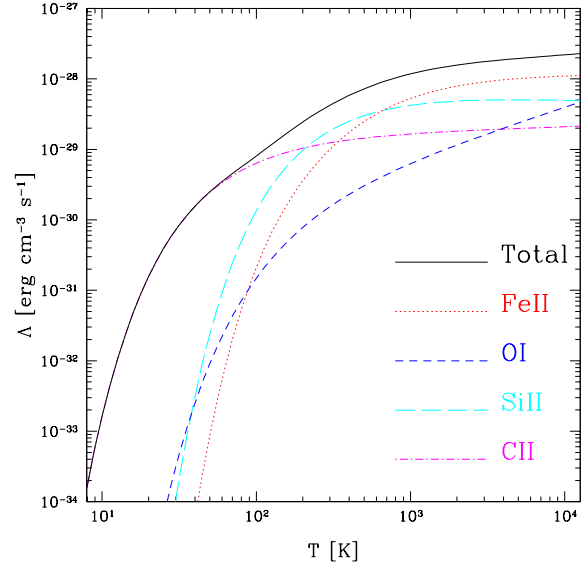}  
\caption{Cooling due to metals as a function of temperature.   
The computations are done for a gas with total number density of $\rm
1\,cm^{-3}$; for each metal species we assume a number density of  
$\rm 10^{-6}\, cm^{-3}$ and we set the free electron over hydrogen  
fraction to a value of $10^{-4}$.
}  
\label{fig:met_line_cooling}  
\end{center}  
\end{figure}  
Combining (\ref{cooling_fct_def}) and (\ref{2lev_sol2})   
one can write the previous equation as a function only of the total  
number density of the species  
  
\begin{equation}\label{2lev_cooling_fct}  
\Lambda\!\! =\!\! \frac{\gamma_{12}^{H} + \gamma_{12}^{e}n_e/n_H  
}{\gamma_{12}^{H}\!\!+\!\!\gamma_{21}^{H}\!\!+\!\!\left(\gamma_{12}^{e}\!\!+\!\!\gamma_{21}^{e}\right)n_e/n_H  
\!\!+\!\! A_{21}/n_H} n_{tot} A_{21} \Delta E_{21}.  
\end{equation}  
For $n_e \ll n_H$, the previous formula is consistent with the one  
quoted in \cite{Santoro_Shull_2006}, who do not consider electron  
impact excitation effects.  
Using equations (\ref{n2_n1}) and (\ref{n_cr}), 
$\Lambda$ can also be written as a  
function of the fundamental level population  
  
\begin{eqnarray}  
\Lambda & = & \frac{ n_1 n_H \gamma_{12}^{H} + n_1 n_e  
\gamma_{12}^{e}} { n_H/n_{cr,21}^{H} + n_{e}/n_{cr,21}^{e} +1  
} \Delta E_{21}\\  
& & \stackrel{n_e\ll n_H}{\sim}  \frac{n_1 n_H  \gamma_{12}^{H}  
}{n_H/n_{cr,21}^{H} + 1 }\Delta E_{21}  
\end{eqnarray}  
being $n_{cr,21}^{H}$ and $n_{cr,21}^{e}$  
 the critical density for the transition $2  
\rightarrow 1$ due to H- and e-impact excitations.\\  
In particular, in the low density limit ($n_{H,e}\ll n_{cr}$), the  
 above equation becomes  
  
\begin{eqnarray}  
\Lambda  \simeq &\left[n_1 n_H \gamma_{12}^{H} + n_1 n_e\gamma_{12}^{e}  
\right] \Delta E_{21} \quad \label{cooling_fct_low_dens1} \\  
 & \stackrel{n_e\ll n_H}{\sim}  n_1 n_H  \gamma_{12}^{H} \Delta E_{21} \label{cooling_fct_low_dens2}.  
\end{eqnarray}  
In this regime, each excitation - see formulae  
(\ref{cooling_fct_low_dens1}) and (\ref{cooling_fct_low_dens2}) - is  
statistically followed by emission of radiation - see the general definition  
(\ref{cooling_fct_def}).\\  
In the high density limit, one finds the expected thermodynamic  
equilibrium cooling rate  
  
\begin{eqnarray}  
\Lambda & \simeq &\frac{g_2}{g_1}e^{-\beta\Delta E_{21}} n_1   
\frac{A_{21}}{1 + n_e\gamma_{21}^{e} /n_H\gamma_{21}^{H} }   
\Delta E_{21} +     \nonumber \\  
&& \frac{g_2}{g_1}e^{-\beta\Delta E_{21}} n_1  
\frac{A_{21}}{1 + n_H\gamma_{21}^{H} /n_e\gamma_{21}^{e} }   
\Delta E_{21} \\  
&\stackrel{n_e \ll n_H}{\sim} &\frac{g_2}{g_1}e^{-\beta\Delta E_{21}} n_1  
A_{21} \Delta E_{21}.\label{high_dens_cooling}  
\end{eqnarray}  
In the right-hand side, it is easy to recognize the Boltzmann  
distribution of populations for $n_2$. It is interesting to note that  
the cooling function does not depend any more on the number density of  
the colliding particles, but only on the species abundance, in  
contrast with the low density regime, where there is a linear  
dependence on both densities.\\  
These arguments ensure that it is safe to use formula   
(\ref{2lev_cooling_fct}) to compute the gas cooling for two-level  
atoms.\\  
  
For $n-$level systems, the cooling function is simply the sum  
of all the contributions from each transition  
  
\begin{equation}\label{cooling_fct_multi}  
\Lambda \equiv \sum_{i\ge 1} \sum_{0\le j < i} n_i A_{ij} \Delta E_{ij} .  
\end{equation}  
In general, once the number density of the cooling species is fixed,  
we expect the cooling function to grow linearly with the colliding  
particle number density and eventually to  
saturate, converging to the Boltzmann statistic, when the critical  
densities are reached.
We see that CII, SiII, FeII saturate when the  
colliding particle number density achieves values around   
$\rm 10^4\,cm^{-3} - 10^5\, cm^{-3}$, while for OI we will have a double  
phase of saturation: the first one at $\rm \sim 10^5\, cm^{-3}$  
involving the lower three states and the second one at $\rm\sim 10^{11}\,
cm^{-3}$ involving the higher two states.\\  
  
As an example, in Figure \ref{fig:met_line_cooling}, we show the  
cooling functions for a total number density $\rm 1\, cm^{-3}$   
and for each metal species $\rm 10^{-6}\,cm^{-3}$;   
the ratio between free electrons and hydrogen is chosen to  
be $\rm 10^{-4}$. With these values, the presence of electrons  
can affect the results up to $10\%$
with respect to the zero electron fraction case.   
We also notice that all the metals contribute with similar importance  
to the total cooling function and the main difference in the cooling  
properties of the gas will depend on their detailed chemical  
composition.\\  
  
\begin{figure}  
\includegraphics[width=0.48\textwidth]{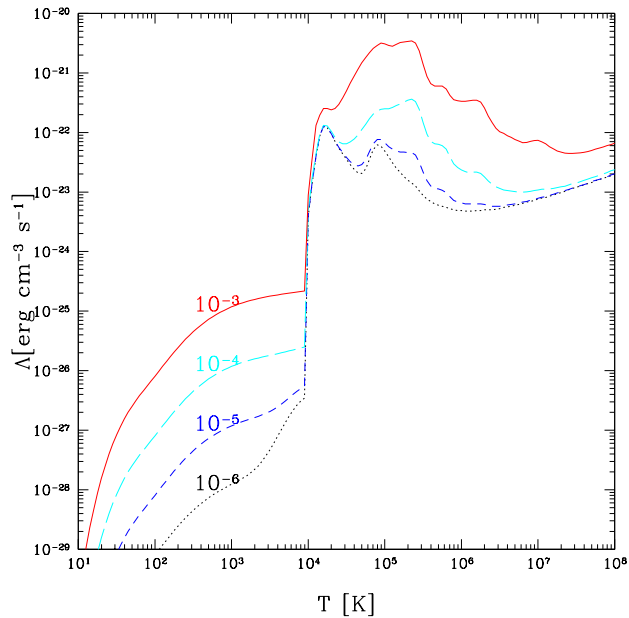}  
\caption{  
Total cooling due to hydrogen, helium, metals, H$_2$ and  
HD molecules as function of temperature, for gas having a hydrogen
number density of 1 $\rm cm^{-3}$.
The fraction of H$_2$ and HD are fixed to   
$10^{-5}$ and $10^{-8}$, respectively.
The labels in the plot refer to different amount of metals, 
for individual metal number fractions of   
$10^{-3}$ (solid line),  
$10^{-4}$ (long-dashed line),  
$10^{-5}$ (short-dashed line) and   
$10^{-6}$ (dotted line).  
}  
\label{fig:cooling_met}  
\end{figure}  

We also plot the cooling functions for all the temperature regime   
we are interested in:
at temperatures higher than $\rm 10^4 ~K$, we interpolate the Sutherland  
and Dopita tables \citep[][]{Sutherland_Dopita_1993}, at lower  
temperatures, we include metals and molecules as discussed previously.
Figure \ref{fig:cooling_met} shows the cooling function for different  
individual metal number fractions with abundances in the range
$10^{-6} - 10^{-3}$ and $\rm H_2$ and HD fractions of $10^{-5}$ and 
$10^{-8}$, respectively. These values for $\rm H_2$ and HD are fairly
typical for the IGM gas at the mean density \citep[see also the
conclusions of][and references therein]{GP98}.
In the temperature range $\rm 10^4 ~K - \rm 10^5 ~K$, the double peak due
to hydrogen and helium collisional excitations is evident at low 
metallicity, while it is washed out by the contribution of different
metal ionization processes as the metallicity increases.
For example, complete collisional ionization of carbon and oxygen produces
the twin peak at $\rm 10^5 ~K$, while complete ionization of iron is
evident at about $\rm 10^7 ~K$.
At temperatures lower than $10^4\rm ~K$ and metal fractions lower than  $\sim  
10^{-6}$, the dominant cooling is given by
molecules; instead, for larger metal fractions the effects of metals became  
dominant.\\  
The general conclusion is  
that at very high redshift, when metals are not present, only $\rm  
H_2$ and HD can be useful to cool the gas down to some $\rm 10^2 ~K$,  
while after the first stars explode, ejecting heavy elements into the  
surrounding medium, metals quickly become the most efficient coolants.


\section{Tests} \label{sect:tests}  
In this section, we are going to test the implementation of HD and  
metal cooling using different kind of simulations. In  
particular, we focus on the analysis of abundance redshift evolution,  
cosmic structure formation and clusters.  

\subsection{Abundance redshift evolution} \label{sect:abundances}  
\begin{figure}  
\begin{center}
\includegraphics[width=0.50\textwidth]{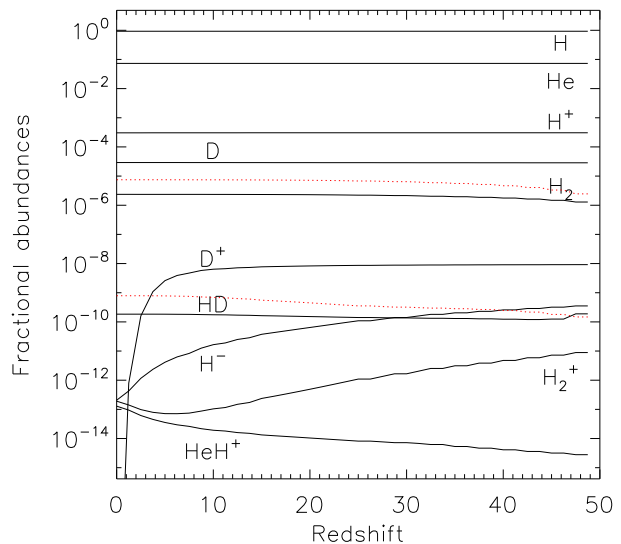}  
\caption{Abundances as a function of redshift. The solid lines refer to  
the abundance evolution in a flat cold dark matter universe   
with   
$h =~ 0.67$,  
$\rm \Omega_{0m} =~ 1$,  
$\rm \Omega_{0b} =~ 0.037$;  
the dotted lines refers to H$_2$   
and HD evolution in a $\Lambda$CDM model with   
$h =~ 0.73$,  
$\rm \Omega_{0m} =~ 0.237$,  
$\rm \Omega_{0\Lambda} =~ 0.763$,  
$\rm \Omega_{0b} =~ 0.041$.  
}  
\label{fig:frac_ab}  
\end{center}  
\end{figure}
As a first test, we investigate the behavior of a plasma of   
primordial chemical composition (i.e. with no metals)   
looking at the redshift evolution of the single abundances.  
Our goal is to reproduce the results from \cite{GP98},   
who calculate the redshift evolution of a metal-free gas at the mean
density by following a detailed chemical network.
For this reason, here, we perform our non-equilibrium computations on  
$isolated$ particles,  
including the following chemical species:  
e$^-$, H, H$^+$, He, He$^+$, He$^{++}$, H$_2$, H$^+_2$, H$^-$,  
D, D$^+$, HD, HeH$^+$  
and assuming a flat cosmology with no dark energy  
content (matter density parameter $\rm\Omega_{0m} = 1$),
baryon density parameter $\rm\Omega_{0b} = 0.037$,
Hubble constant, in units of $\rm 100 km\, s^{-1}\, Mpc^{-1}$,   
$h = 0.67$ and initial gas temperature of $1000 \rm ~K$.

The evolution of the number fractions for the different species is  
plotted in Figure \ref{fig:frac_ab};  
the electron abundance is given from charge conservation of neutral  
plasma and is normally very close to the $\rm H^{+}$ value, this being  
the dominant ion.
These results are in very good agreement with those of \cite{GP98}.\\  
  
In our set of reactions, due to the low initial gas temperature,  
the collisions are inefficient to ionize helium.  
The inclusion of $\rm HeH^+$ creation  
  
\begin{equation}  
\rm He + H^+  \rightarrow HeH^+  + \gamma  
\end{equation}  
contributes to rise H$_2^{+}$ abundance mainly via reaction   
  
\begin{equation}  
\rm HeH^+ + H \rightarrow He + H_2^+  
\end{equation}  
and weakly decrease the $\rm H^{-}$ number fraction via   
  
\begin{equation}  
\rm HeH^+ + \gamma  \rightarrow He + H^+  
\end{equation}  
  
\begin{equation}
\rm H^+ + H^-  \rightarrow  2H
\end{equation}  
where $\gamma$ indicates the photons.
Because of the very low $\rm HeH^+$ abundance reached,  
there is no substantial  $\rm He$ atom abundance evolution.

Another $caveat$ to take into account is the lack of reactions between  
$\rm D^+$ and free electrons which would destroy the deuterium ions more  
efficiently, but without altering significantly the global amount of  
HD formed.  
We notice also the exponential decay of $\rm D^{+}$ due to the rate
coefficient of equation (\ref{kHD5}) and the freezing out of   
$\rm H^{+}$, $\rm H_2$, D and HD number fractions.
  
As a comparison, we also plot (dotted lines) 
the H$_2$ and HD abundance evolution in a  
flat $\Lambda$CDM model having
$h =~ 0.73$,   
$\rm \Omega_{0m} = 0.237$,   
$\rm \Omega_{0\Lambda} = 0.763$,   
$\rm \Omega_{0b} = 0.041$ (Spergel et al. 2006).   
The slight increment observed is due to the fact that in the cold dark  
matter cosmology   
the baryon fraction is about $4\%$, making the interactions among  
different species rarer than in the $\Lambda$CDM model, for which the  
baryon fraction is about $17\%$. In addition, the cosmological constant is   
dominant only at redshifts below one.
The evolution of the other species is similar in both cosmologies.\\  

\subsection{Cosmic structure formation} \label{sect:large_scale}
To test the behavior of the code in simulations of structure
formation and evolution and the impact of HD, we run a cosmological
simulation with the same properties and cosmology as in
\cite{UMaio_et_al_2006}.
The main difference here is the addition of HD chemistry.
We adopt the concordance $\rm \Lambda CDM$ model with
$h = 0.7$,
$\rm \Omega_{0m} = 0.3$,
$\rm \Omega_{0b} = 0.04$,
$\rm \Omega_{0\Lambda}= 0.7$;
the power spectrum is normalized assuming a mass variance in a   
$\rm 8\, Mpc/h$ radius sphere $\sigma_8 = 0.9$ and the spectral index  
is chosen to be $\rm n = 1$.
We sample the cosmological field (in a periodic box of 
1 Mpc comoving side length) with $324^3$
dark matter particles and the same number of gas particles, having a  
mass of about $\rm 1040\,M_{\odot}$ and $\rm 160\,M_{\odot}$,
respectively.
 The comoving Plummer-equivalent gravitational softening length is  
fixed to $\rm 0.143\,kpc$. This allows to resolve halos with mass of
about $10^5 M_{\odot}$. The simulation starts at $z=90$ and is
stopped at $z \sim$21.\\
We include the reactions involving e$^-$, H,
H$^+$, He, He$^+$, He$^{++}$, H$_2$, H$^+_2$, H$^-$, D, D$^+$, HD 
(here we neglect $\rm HeH^+$, as it has not significant effects on the
simulation)
and compare the results with those of \cite{UMaio_et_al_2006}, whose
$\Lambda$CDM simulation has the same features, but the chemical set
does not follow the evolution of D, D$^+$ and HD and does not include $\rm H_2^+$ cooling.\\

\begin{figure}  
\begin{center}  
\includegraphics[width=0.45\textwidth]{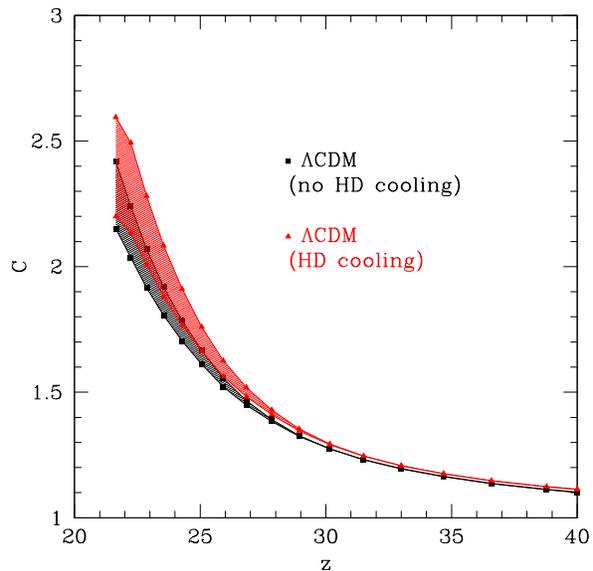}  
\caption{Gas clumping factor as a function of redshift for two  
$\rm \Lambda$CDM models with different chemical composition.   
The squares refer to the clumping factor computed   
with standard atomic line cooling and $\rm H_2$ cooling,  
while the triangles refer to a case which includes also HD cooling. 
The shaded regions correspond to the variation of the maximum overdensity between 100 (lower line in both cases) and 500 (upper line in both cases).  
}
\label{fig:clumping}  
\end{center}
\end{figure}
To quantify the differences between the two runs and the efficiency of the   
HD cooling we calculate the gas clumping factor, $C$, in the
simulation box, in the following way
\begin{equation}  
C = \frac{\sum_i m_i\rho_i \sum_j m_j\rho_j^{-1}}{\left(\sum_k m_k\right)^2}  
\end{equation}
where for each SPH particle, $i$, we indicate with  
$m_i$ its mass and with $\rho_i$ its mass density; the indices   
run over all the gas particles.  For the sake of comparison,
we calculate $C$ using only particles  
with density below a given overdensity threshold, $\delta_M$, and we  
make $\delta_M$ vary in the range [100,~500].\\
The results are plotted in Figure \ref{fig:clumping} for both  
simulations.
We see that the inclusion of HD makes the clumping factor increase  
at all redshifts, almost independently from the density threshold. This  
means that the gas is, on average, denser and more  
clumped, with an increment of about 10\% at redshift $\sim$22.
  
\subsection{Cluster} \label{sect:cluster}  

So far, we have assumed either primordial gas with no metal pollution  
(previous test case) or a pre-defined metallicity to demonstrate the  
effect of the presence of metals on the cooling function at low temperature, 
as in Section \ref{sect:cooling}.
Now, we are going to couple our cooling function with a model for 
the chemical enrichment and test this implementation within a simulation
that follows the formation of a cluster. 
In addition to testing the validity of our implementation,
although there are no significant changes for the intra-cluster medium (ICM)
to be expected, it is of interest to check whether there are regions
inside the simulations where the polluted medium is cooling below $\rm
10^4 ~K$ due to its metal content.\\
The ``zoomed initial condition technique''
\citep[][]{Tormen_et_al_1997} is used to extract 
from a dark matter-only simulation with box size of $\rm  
479\,Mpc/h$ (we adopt a $\rm \Lambda CDM$ cosmology with
 $\rm H_0 =~ 70\, km\,s^{-1}\, Mpc^{-1}$, 
$\sigma_8 =~ 0.9$, $\rm \Omega_{0\Lambda}=~ 0.7$,
$\rm \Omega_{0m} =~ 0.3$, $\rm \Omega_{0b} =~ 0.04$) a smaller region and
to re-simulated it at higher resolution introducing also gas particles.
The cluster evolution is simulated with about $2\cdot 10^5$ particles.
The comoving Plummer-equivalent gravitational softening length is
$\rm 5 \, kpc/h$.
At redshift zero, the selected cluster has a virial mass of about 
$\rm 10^{14} M_{\odot}/h$, a virial radius of about
$\rm 1 Mpc/h$ and a virial temperature of $2\cdot 10^7$ ~K 
 \citep[for more details see][]{Dolag2004}.\\
We start the simulation with no metallicity content. Then, the metal
abundances are consistently derived \citep[as in][]{Tornatore_et_al_2007}
following the star formation history of the system, accounting for the 
lifetime of stars of different mass \citep[][]{PM1993} distributed according
to a Salpeter IMF and adopting appropriated stellar yields:   
we use those from
\cite{WW1995} for massive stars (SNII),  
\cite{vdH_G_1997} for low- and intermediate-mass stars and
\cite{Thielemann_et_al_2003} for SNIa.  
The underlying sub-resolution model for star 
formation in multi-phase interstellar medium  
\citep[][]{SpringelHernquist03} includes a 
phenomenological model for feedback from galactic ejecta powered by  
the SNII explosions, where we have chosen the wind velocity  
to be $\rm 480 km\,s^{-1}$.
As we are only interested to test the effect of the metals, 
we exclude H$_2$, HD and $\rm HeH^+$ chemistry and consider only  
atomic cooling from collisional excitations of hydrogen and helium.
Once the medium gets polluted with metals, their contribution  
is added. For the metal cooling of the gas above $10^4$~K,  
Sutherland and Dopita tables \citep[][]{Sutherland_Dopita_1993} are  
used. At lower temperatures, the fine structure transitions from OI, 
CII, SiII, FeII are included as discussed in the previous Sections.  

\begin{figure}
\begin{center}
\includegraphics[width=0.45\textwidth]{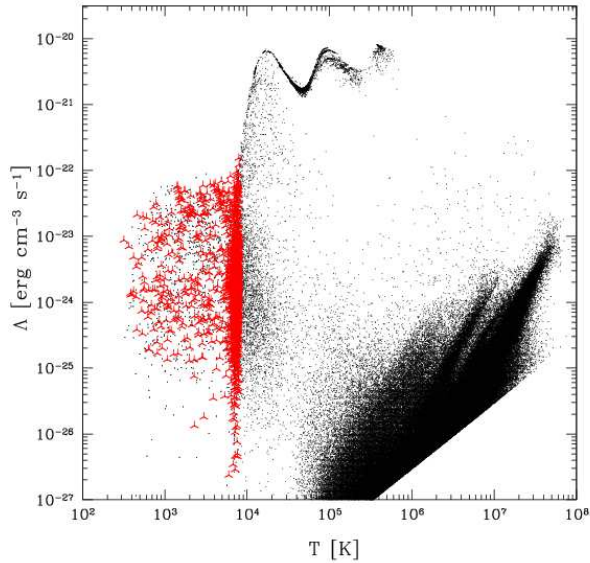}  
\caption{
Distribution of the particles of a cluster simulation in the  
T-$\Lambda$ space. The hot and thin intra-cluster medium  
populates the bottom right area. The particles within collapsed
objects, which represent very dense regions of the simulation,  
populate the high temperature cooling function (upper branch).  
In addition, metal enriched particles undergo metal 
line cooling in the low temperature regime (below $\sim10^4$ ~K).  
The three-pointed star symbols correspond to particles which are  
located within twice the virial radius of the cluster and have a  
temperature lower than $\rm 8000 ~K$.
}
\label{fig:TL}
\end{center}
\end{figure}
\begin{figure}
\begin{center}
\includegraphics[width=0.45\textwidth]{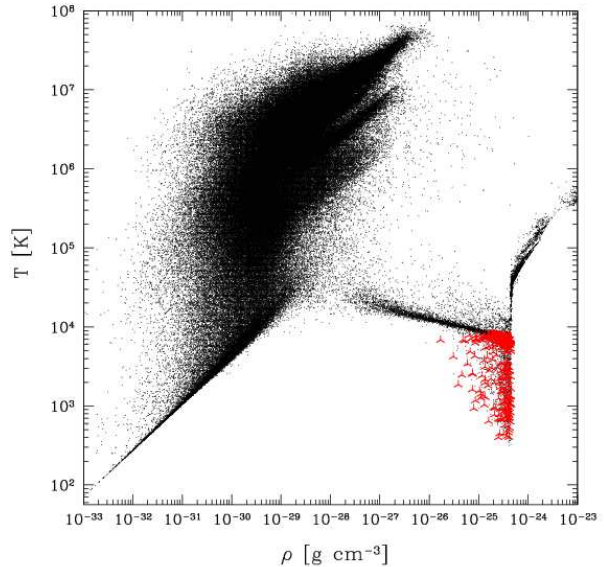}
\caption{
Distribution of the particles of a cluster simulation in the phase
diagram. The hot and thin intra-cluster medium populates the
central-left area of the plot, while the dense and cooled regions of the
simulation are represented in the lower-right part. Particles heated
by feedback effects are represented by points in the central-right
side. The three-pointed star symbols correspond to metal
enriched particles which are located within twice the virial radius of
the cluster and have a temperature lower than $\rm 8000 ~K$.
We remind that the critical density of the Universe at the present is
$\rho_{0,cr} \simeq 1.9\cdot 10^{-29}h^2 \quad
g\,cm^{-3}$.
}
\label{fig:rhoT}
\end{center}
\end{figure}

Figure \ref{fig:TL} shows the cooling diagram of our simulation  
at redshift $z=0$; each SPH particle is represented by a point.
In the plot, different areas can be identified.
The one at high temperatures (bottom right) represents the hot ICM.  
When the ICM starts to get denser, cooling gets more efficient:
the corresponding gas particles are represented by the points belonging to 
the upper branch of the cooling function  
and they are  brought to lower and lower temperatures.
Feedback from the star
formation partially pushes some of them away from the cooling curve to  
slightly higher temperatures. Below $\rm 10^4 ~K$, only particles which  
are metal enriched can further cool down to about $\rm 300 ~K$, while  
gas particles with primordial composition are stacked at $\rm 10^4 ~K$.
The three-pointed stars refer to particles within twice the  
virial radius and with a temperature below $\rm 8000~K$, indicating that  
this region of the T-$\Lambda$ space is also populated by gas  
associated with the galaxy cluster.\\
The corresponding phase diagram is shown in Figure \ref{fig:rhoT}.
The three-pointed star symbols are the same as in Figure \ref{fig:TL}.
The hot and thin intra-cluster medium populates the
central-left area of the plot, while the dense and cool regions 
occupy the lower-right part. Particles heated
by feedback are represented by points in the central-right side
($\rm \rho >~ 10^{-24} \, g\, cm^{-3}$, $\rm T >~ 10^5\,K$~).
The main effect of our metal cooling implementation is to lower the
temperature of the dense medium, generating the sharp triangular area
visible in the $\rm \rho~$-T space, at $\rm T < 10^4 ~K$ and $\rm
\rho~>~10^{-26} \,g\, cm^{-3}$.
The points at very low densities are associated with diffuse metal
free gas; this suggests that the spread in the cooling diagram of
Figure \ref{fig:TL} at
temperature lower than $\rm 10^4 ~K$ is mainly due to different fractional
metal enrichment of the particles, rather then to their different densities.\\
As already mentioned, global properties of the ICM and star formation
are not significantly changed compared to the reference run without
the metal line cooling from fine structure transitions.
This happens because the simulation
was merely meant to be a test case of the implementation of metal line
cooling below $\rm 10^4 ~K$ under realistic conditions, but the halos resolved
are large enough to cool and form stars without the aid of such
cooling. In order
to investigate in more detail the effects of the additional cooling by 
molecules and metals at low temperatures on the ICM and the star formation,
higher resolution simulations are needed. 
However, this opens interesting grounds for further investigations on
the interplay between formation of small objects, with virial
temperatures in the range of interest for our extended cooling
function, and metal pollution from first stars.

\section{Conclusions}\label{sect:conclusions}  
In order to understand structure formation and evolution,   
a detailed study of the chemical and cooling properties of baryonic matter  
is needed.\\
In this paper, we have presented time dependent calculations of the  
cooling properties of a gas in a ``low temperature'' regime, using the  
contributions of several chemical species and we have tested the effect on  
cosmic structure evolution.\\
Hydrogen derived molecules are
effective in cooling metal-free gas below a temperature of $\rm\sim
10^4 ~K$, the typical temperature range of primordial objects. On the  
other hand, when the medium is polluted by material expelled from  
stars (via SN explosions, mass losses in AGB phase and winds),  
metals are expected to become the main coolants.\\  
For these reasons, we have extended previous ``non-equilibrium''  
calculations \citep[][]{Yoshida_et_al2003}
in order to include in the numerical code Gadget-2
\citep{Springel2001, Springel2005},
the deuterium chemistry and follow the  
formation/destruction of HD molecule. This, together with molecular hydrogen,  
is able to cool down the gas at $\rm T ~<~ 10^4 ~K$.   
Thanks to its permanent electric dipole moment, HD could allow cooling  
even below $\rm 10^2 ~K$ \citep[][]{Yoshida2006_astroph}.\\
Other molecules are not very significant for the gas cooling properties.\\  
The treatment of metal cooling at $\rm T \ge 10^4~K$ is included using  
the tables provided by \cite{Sutherland_Dopita_1993}, while the  
contribution from fine structure transitions of oxygen, carbon,  
silicon and iron at $T< 10^4$~K has been included 
by computing the populations of the levels, for each species, using
the detailed balancing principle.
More in particular, we have assumed that  
the UV radiation coming from the parent star ionizes carbon, silicon  
and iron, while oxygen remains neutral as its first ionization  
potential is higher than $\rm 13.6~eV$. We deal with the gas radiative  
losses computing the detailed balancing populations of the levels due  
to collisional excitations arising from hydrogen and electron  
impacts.  The cooling follows the level de-excitations.  
The electron impact excitations are also included,   
as a residual electron fraction of about $10^{-4}$ survives in  
the post-recombination epoch and higher values are reached during the  
reionization process.\\  
On the whole, we are now able to follow the evolution of  
e$^-$, H, H$^+$, He, He$^+$, He$^{++}$, H$_2$, H$^+_2$, H$^-$,  
D, D$^+$, HD, $\rm HeH^+$, O, C$^+$, Si$^+$, Fe$^+$; so, the code is  
suitable to deal both with primordial and metal enriched gas.\\  
We have checked the validity of our scheme by comparing
the results of some 
test runs with previous calculations of molecule abundance evolution,  
finding excellent agreement.\\  
We have also investigated the relevance of HD and metal cooling in  
some specific cases.\\  
Adding the deuterium chemistry and HD contribution to the cooling function  
in simulations of structure formation results in a higher clumping  
factor of the gas, i.e. clouds are slightly denser and more compact,  
at high redshifts, with respect to the case  
when only H, He and $\rm H_2$ cooling is considered.   
The difference is about $10\%$ at $\rm z\sim 22$.\\  
For what concerns the role of metal cooling at $\rm T<10^4 ~K$, we have  
shown that their presence is relevant in this temperature regime. In  
particular, in the cluster simulations we have run, fine structure  
transitions can actually cool the local temperature down to $\rm 10^2~K  
- 10^3~K $.\\  
  
In conclusion, we have implemented in Gadget-2, the most relevant
features of 
gas cooling, in both pristine and polluted environments, for the 
temperature range $\rm 2.7 ~K - 10^9 ~K$.  
We find that HD cooling has some influence  
on the high redshift gas clumping properties,
while low temperature metal cooling has a significant impact on the  
formation and evolution of cold objects.
In addition to investigating the above topics, this implementation can be used 
to study the detailed enrichment history of the IGM and its possible 
interplay with the transition between a primordial, massive star formation
mode and a more standard one.


\section*{acknowledgements}
We acknowledge profitable discussions with M.~Ricotti and 
useful comments from G.~De~Lucia and the anonymous referee;
U.~M. is thankful to N.~Yoshida for his helpful suggestions.\\  
Computations were performed on the machines at the
computing center of the Max Planck Society with CPU time assigned to  
the Max Planck Institute for Astrophysics.  
  
  
\begin{appendix}  
  
\section{Chemical rates}\label{app1}  
  
We consider the following set of  
equations involving HD creation and destruction:  
\begin{eqnarray}  
\rm	 H_2\phantom{\,}  +  D\phantom{^+}    &\rightarrow &\rm HD + H 	\\  
\rm 	 H_2\phantom{\,}  +  D^+ &\rightarrow &\rm HD + H^+ \\  
\rm	HD + H\phantom{^+}    &\rightarrow &\rm D + H_2 	 \\  
\rm	HD + H^+  &\rightarrow &\rm D^+ + H_2 	\\  
\rm	H^+ + D\phantom{^+}  & \rightarrow &\rm H + D^+		 \\  
\rm	H\phantom{^+} + D^+  & \rightarrow &\rm H^+ + D.	  
\end{eqnarray}  
We use the rate coefficients from \cite{Wang_Stancil_2002}:  
\begin{eqnarray}  
	k_{\rm HD,1} & = & 9.0\cdot 10^{-11}  e^{-3876/T} \rm\, cm^3\, s^{-1} \\  
	k_{\rm HD,2} & = & 1.6 \cdot 10^{-9} \rm cm^3 s^{-1}   
\end{eqnarray}  
\cite{SLD_1998}:  
\begin{eqnarray}  
      	k_{\rm HD,3} & = & 3.2 \cdot 10^{-11} e^{-3624/T} \rm \,cm^3 \,s^{-1} \\  
      	k_{\rm HD,4} & = & 10^{-9} e^{-464/T} \rm cm^3 s^{-1}   
\end{eqnarray}  
and \cite{Savin_2002}:  
\begin{eqnarray}  
      k_{\rm HD,5} &\!\!\!\!\!\!=\!\!\!\!\!\! & 2 \!\!\cdot\!\! 10^{-10}  
T^{0.402}e^{-37.1/T} \!\!-\!\!  
	3.31 \!\!\cdot\!\! 10^{-17} T^{1.48} \rm cm^3 s^{-1} \\  
      k_{\rm HD,6} &\!\!\!\!\!\!\!\!=\!\!\!\!\!\!\!\! & 2.06 \!\!\cdot\!\!  
10^{-10} T^{0.396}\!e^{-33.0/T}  
	\!\!\!\!+\!\! 2.03\!\!\cdot\!\! 10^{-9} T^{-0.332} \!\!\rm \,cm^3 \,s^{-1}.  
\end{eqnarray}  
\\
We consider the main equations for $\rm HeH^+$ formation and  
evolution \citep{GP98}, namely:  
\begin{eqnarray}  
\rm	He\phantom{-}+  H^+ & \rightarrow   &\rm HeH^+ + \gamma\\  
\rm	HeH^+ + H           & \rightarrow   &\rm He  + H_2^+\\  
\rm	HeH^+ + \gamma      & \rightarrow   &\rm He + H^+   
\end{eqnarray}  
and the rates from \cite{RD1982}:   
\begin{equation}  
k_{\rm HeH^{+}, 1} =   
	\left\{   
      	\begin{array}{ll}  
		\rm 7.6 \cdot 10^{-18} T^{-0.5} \,cm^3\, s^{-1}, &\rm T\leq 10^3~K\\ 
		&       \\	  
		\rm 3.45\cdot 10^{-16} T^{-1.06}\, cm^3\, s^{-1}, &\rm T > 10^3~K\\  
     	\end{array}  
     	\right.  
\end{equation}  
\cite{KAH1979}:  
\begin{eqnarray}  
k_{\rm HeH^{+}, 2} = 9.1 \cdot 10^{-10} \rm \,cm^3 \,s^{-1}  
\end{eqnarray}  
and \cite{RD1982}:  
\begin{eqnarray}  
k_{\rm HeH^{+}, 3} = 6.8\cdot 10^{-1} T_r^{1.5}e^{-22750/T_r}  \rm\,s^{-1}.  
\end{eqnarray}  
\\  
In the previous expressions, $T$ stands for the gas  
temperature and $T_r$ for the radiation temperature.  
  

\section{Atomic data}\label{app2}
In the following, the atomic data adopted in this paper are provided  
\citep[see also][]{Osterbrock1989, HM89, Santoro_Shull_2006}.  
We will use the usual spectroscopic notation for many electron atoms:   
 {\bf S} is the total electronic spin quantum operator,   
{\bf L}  the total electronic orbital angular momentum operator and 
{\bf J}~=~{\bf L}~+~{\bf S} the sum operator; 
S, L, J, are the respective quantum numbers  
and X indicates the orbitals S, P, D, F, ...,  
according to L=0, 1, 2, 3, ..., respectively; then $\phantom{}^{2S+1}X_J$  
will indicate the atomic orbital X, with spin quantum number S and  
total angular momentum quantum number J;  
its multiplicity is equal to $2J+1$.\\
In the following, we are going to discuss the models adopted for each  
species and the lines considered in a more detailed way.\\  
We will often use the notation $\rm T_{100}=T/100 ~K$.\\

\begin{itemize}  
  
\item[*]
{\bf CII}: we model CII as a two-level system considering the fine structure transition $(2p)[\phantom{}^2P_{3/2}-\phantom{}^2P_{1/2}]$ between the quantum  
number $J=3/2$ and $J=1/2$ states.  
Data were taken from \cite{HM89}:\\  
$\phantom{}$\\  
$\gamma_{21}^{H} = 8\cdot 10^{-10} T_{100}^{0.07} \quad\rm cm^3\, s^{-1}$;\\  
$\gamma_{21}^{e} = 2.8\cdot 10^{-7} T_{100}^{-0.5}\quad\rm cm^3\, s^{-1}$;\\  
$A_{21} = 2.4\cdot 10^{-6} \quad\rm s^{-1}$;\\  
$\Delta E_{21} = 1.259\cdot 10^{-14} \rm \,erg$  
$\quad\rm(T_{exc} = 91.2 ~K, \lambda = 157.74\,\mu m)$.\\
$\phantom{}$\\  
The cooling function is computed according to formula (\ref{2lev_cooling_fct}).\\  
  
\item[*]   
{\bf SiII}: we model SiII as a two-level system with the fine structure  
transition $(3p)[\phantom{}^2P_{3/2}-\phantom{}^2P_{1/2}]$. Data  
were taken from \cite{HM89}:\\  
$\phantom{}$\\  
$\gamma_{21}^{H} = 8\cdot 10^{-10} T_{100}^{-0.07} \quad\rm cm^3\,s^{-1}$;\\  
$\gamma_{21}^{e} =  1.7\cdot 10^{-6} T_{100}^{-0.5} \quad\rm cm^3\,s^{-1}$;\\  
$A_{21} =  2.1\cdot 10^{-4} \quad\rm s^{-1}$;\\  
$\Delta E_{21} = 5.71\cdot 10^{-14}\, \rm \,erg$  
$\quad\rm(T_{exc} = 413.6 ~K, \lambda = 34.8\,\mu m)$.\\  
$\phantom{}$\\  
The cooling function is computed following formula (\ref{2lev_cooling_fct}).\\  
  
\item[*]   
{\bf OI}: neutral oxygen is a metastable system formed by the (S=1, L=1) triplet and  
(S=0, L=0,2) doublet, $(2p)[\phantom{}^3P_2 - \phantom{}^3P_1 -  
\phantom{}^3P_0 - \phantom{}^1D_2 - \phantom{}^1S_0]$, in order of  
increasing level, with the following excitations rates   
\citep[][]{HM89, Santoro_Shull_2006}:\\  
$\phantom{}$\\  
$\gamma^{H}_{21} = 9.2\cdot 10^{-11} T_{100}^{0.67} \quad\rm cm^3\,s^{-1}$;\\  
$\gamma^{H}_{31} = 4.3\cdot 10^{-11} T_{100}^{0.80} \quad\rm cm^3\,s^{-1}$;\\  
$\gamma^{H}_{32} = 1.1\cdot 10^{-10} T_{100}^{0.44} \quad\rm cm^3\,s^{-1}$;\\  
$\gamma^{H}_{41} = \gamma^{H}_{42}= \gamma^{H}_{43}= 10^{-12}\quad\rm cm^3\,s^{-1}$;\\  
$\gamma^{H}_{51} = \gamma^{H}_{52}= \gamma^{H}_{53}= 10^{-12}\quad\rm cm^3\,s^{-1}$;\\
$ \gamma^{e}_{21} = 1.4\cdot 10^{-8}\quad\rm cm^3\,s^{-1}$;\\  
$ \gamma^{e}_{31} = 1.4\cdot 10^{-8}\quad\rm cm^3\,s^{-1}$;\\  
$ \gamma^{e}_{32} = 5.0\cdot 10^{-9}\quad\rm cm^3\,s^{-1}$;\\  
$ \gamma^{e}_{41} = \gamma^{e}_{42} = \gamma^{e}_{43} = 10^{-10}\quad\rm cm^3\,s^{-1}$;\\  
$ \gamma^{e}_{51} = \gamma^{e}_{52} = \gamma^{e}_{53} =  
10^{-10}\quad\rm cm^3\,s^{-1}$.\\  
$\phantom{}$\\
The radiative transition probabilities are   
\citep{Osterbrock1989, HM89}:\\  
$\phantom{}$\\  
 $ A_{21} = 8.9 \cdot 10^{-5} \quad\rm s^{-1}$; \\
 $ A_{31} = 1.0 \cdot 10^{-10} \quad\rm s^{-1}$;\\  
 $ A_{32}  = 1.7 \cdot 10^{-5} \quad\rm s^{-1}$;\\  
 $ A_{41}  = 6.3 \cdot 10^{-3} \quad\rm s^{-1}$;\\  
 $ A_{42}  = 2.1 \cdot 10^{-3} \quad\rm s^{-1}$;\\  
 $ A_{43}  = 7.3 \cdot 10^{-7} \quad\rm s^{-1}$;\\  
 $ A_{51}  = 2.9 \cdot 10^{-4} \quad\rm s^{-1}$;\\  
 $ A_{52}  = 7.3 \cdot 10^{-2} \quad\rm s^{-1}$;\\  
 $ A_{54}  = 1.2 \quad\rm s^{-1}$;\\  
$\phantom{}$\\  
energy separations are derived from \cite{HM89}:\\  
$\phantom{}$\\  
$\Delta E_{21} = 3.144\cdot 10^{-14}\,\rm erg$  
$\quad\rm(T_{exc} = 227.7 ~K, \lambda = 63.18\,\mu m)$;\\  
$\Delta E_{32} = 1.365\cdot 10^{-14}\,\rm erg$  
$\quad\rm(T_{exc} = 98.8 ~K, \lambda = 145.5\,\mu m)$;\\  
$\Delta E_{43} = 3.14\cdot 10^{-12}\,\rm erg$  
$\quad\rm(T_{exc} =  2.283\cdot 10^{4}~K, \lambda = 6300\,\AA)$;\\  
$\Delta E_{53} = 3.56\cdot 10^{-12}\, \rm erg$  
$\quad\rm(T_{exc} = 2.578\cdot 10^{4}~K, \lambda = 5577\,\AA)$.\\  
$\phantom{}$\\  
To compute the cooling function, we solve for the five level populations  
and sum over the contributions from each of them.\\

\item[*]   
{\bf FeII}: we adopt a model for a five-level system including  
the transitions   
$(3d) [\phantom{}^6D_{9/2} - \phantom{}^6D_{7/2} - \phantom{}^6D_{5/2}  
-\phantom{}^6D_{3/2} - \phantom{}^6D_{1/2}]$ in order of increasing level.  
For the data see also \cite{Santoro_Shull_2006} and references therein:\\  
$\phantom{}$\\  
$  \gamma^{H}_{21} = 9.5 \cdot 10^{-10}\quad\rm cm^3\, s^{-1}$;\\  
$  \gamma^{H}_{32} = 4.7 \cdot 10^{-10}\quad\rm cm^3\, s^{-1}$;\\  
$  \gamma^{H}_{43} = 5. \cdot 10^{-10}\quad\rm cm^3\, s^{-1}$;\\  
$  \gamma^{H}_{54} = 5 \cdot 10^{-10}\quad\rm cm^3\, s^{-1}$;\\
$  \gamma^{H}_{31} = 5.7 \cdot 10^{-10}\quad\rm cm^3\, s^{-1}$;\\  
$  \gamma^{H}_{41} = 5 \cdot 10^{-10}\quad\rm cm^3\, s^{-1}$;\\
$  \gamma^{H}_{51} = 5 \cdot 10^{-10}\quad\rm cm^3\, s^{-1}$;\\
$  \gamma^{H}_{42} = 5 \cdot 10^{-10}\quad\rm cm^3\, s^{-1}$;\\
$  \gamma^{H}_{52} = 5 \cdot 10^{-10}\quad\rm cm^3\, s^{-1}$;\\
$  \gamma^{H}_{53} = 5 \cdot 10^{-10}\quad\rm cm^3\, s^{-1}$;\\
$ \gamma^{e}_{21} = 1.8\cdot 10^{-6} T_{100}^{-0.5}\quad\rm cm^3\, s^{-1}$;\\  
$ \gamma^{e}_{32} = 8.7\cdot 10^{-7}T_{100}^{-0.5}\quad\rm cm^3\, s^{-1}$;\\  
$ \gamma^{e}_{43} = 10^{-5}T^{-0.5}\quad\rm cm^3\, s^{-1}$;\\
$ \gamma^{e}_{54} = 10^{-5}T^{-0.5}\quad\rm cm^3\, s^{-1}$;\\
$ \gamma^{e}_{31} = 1.8\cdot 10^{-6}T_{100}^{-0.5}\quad\rm cm^3\, s^{-1}$;\\  
$ \gamma^{e}_{41} = 10^{-5}T^{-0.5}\quad\rm cm^3\, s^{-1}$;\\
$ \gamma^{e}_{51} = 10^{-5}T^{-0.5}\quad\rm cm^3\, s^{-1}$;\\
$ \gamma^{e}_{42} = 10^{-5}T^{-0.5}\quad\rm cm^3\, s^{-1}$;\\
$ \gamma^{e}_{52} = 10^{-5}T^{-0.5}\quad\rm cm^3\, s^{-1}$;\\
$ \gamma^{e}_{53} = 10^{-5}T^{-0.5}\quad\rm cm^3\, s^{-1}$;\\
$\phantom{}$\\  
we assume a fiducial normalization of $10^{-5}$ for missing data  
on e-impact rates. We have checked that the level populations are  
almost insensitive to the adopted values.\\  
$\phantom{}$\\
$  A_{21} = 2.13 \cdot 10^{-3} \quad\rm s^{-1}$;\\  
$  A_{32} = 1.57 \cdot 10^{-3} \quad\rm s^{-1}$;\\  
$  A_{31} = 1.50 \cdot 10^{-9} \quad\rm s^{-1}$;\\  
$  A_{43} = 7.18 \cdot 10^{-4} \quad\rm s^{-1}$;\\  
$  A_{54} = 1.88 \cdot 10^{-4} \quad\rm s^{-1}$;\\  
$\phantom{}$\\  
$  \Delta E_{21} = 7.64 \cdot 10^{-14} \rm \,erg$  
$\quad\rm(T_{exc} = 553.58 ~K, \lambda = 25.99\,\mu m)$;\\  
$  \Delta E_{32} = 5.62 \cdot 10^{-14} \rm \,erg$  
$\quad\rm(T_{exc} = 407.01 ~K, \lambda = 35.35\,\mu m)$;\\  
$  \Delta E_{43} = 3.87 \cdot 10^{-14} \rm \,erg$  
$\quad\rm(T_{exc} = 280.57~K, \lambda = 51.28\,\mu m)$;\\  
$  \Delta E_{54} = 2.27 \cdot 10^{-14} \rm \,erg$  
$\quad\rm(T_{exc} = 164.60~K, \lambda =  87.41\,\mu m)$.\\  
$\phantom{}$\\
To compute the cooling function, we solve for the five   
level populations and sum over the contributions from each of them.\\  
\end{itemize}

\begin{figure}  
\begin{center}  
\includegraphics[width=0.45\textwidth]{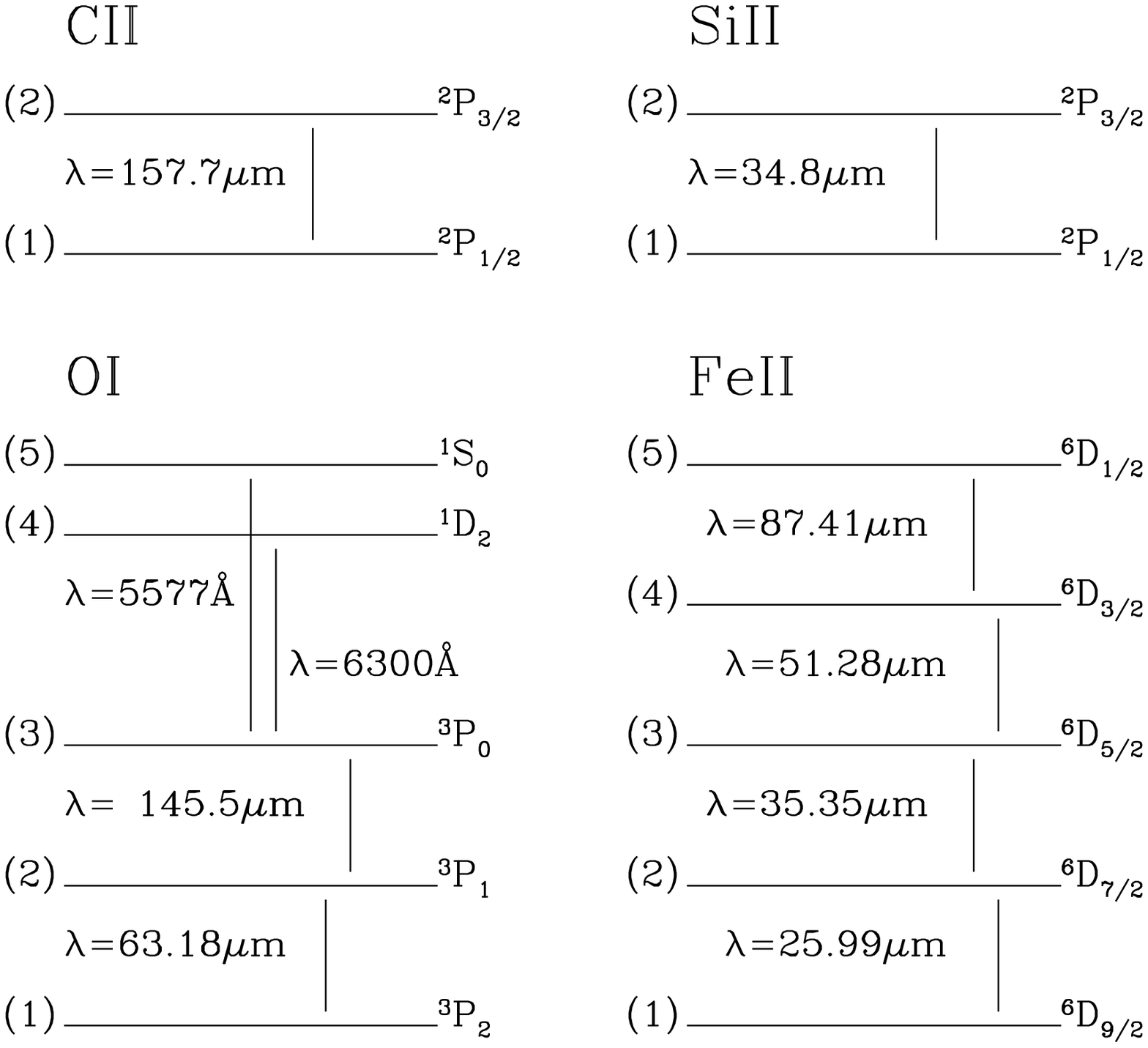}  
\caption{Scheme of the level models adopted for the different atoms  
with respective line transition data.}  
\label{fig:levels}  
\end{center}  
\end{figure}

A scheme of the atomic states, with wavelengths of the transitions
between different levels, is given in Figure \ref{fig:levels}.\\

\end{appendix}

\bibliographystyle{mn2e}
\bibliography{bibl.bib}
\label{lastpage}

\end{document}